\documentclass[fleqn,10pt]{wlscirep}
\usepackage[utf8]{inputenc}
\usepackage[T1]{fontenc}
\usepackage{amsmath,amssymb}
\usepackage{colortbl}
\usepackage{tabulary}
\usepackage{multirow}
\usepackage{wasysym}
\usepackage{array}
\usepackage{xcolor}
\usepackage{tikz}
\usepackage{soul}
\usepackage{dblfloatfix} 
\usepackage[switch]{lineno}

\usetikzlibrary{positioning,shapes.multipart,calc,arrows.meta}
\tikzset{
  basic/.style={draw, text centered},
  rect/.style={basic, text width=2.5em, text height=1.5em, text depth=.5em,fill=blue!20, text centered, rounded corners,line width=1pt},
  1 up 1 down/.style={basic, text width=1.5em, rectangle split, rectangle split horizontal=false, rectangle split parts=2},
  towards/.style={->,line width=1pt},
  from/.style={<-,line width=1pt},
}

\usepackage{makecell}
\setcellgapes{0.01pt}

\title{Turnover in close friendships: age and gender differences}

\author[1,*]{Chandreyee Roy}
\author[1,2]{Kunal Bhattacharya}
\author[3]{Robin I.M. Dunbar}
\author[1,4]{Kimmo Kaski}
\affil[1]{Department of Computer Science, Aalto University School of Science, Espoo, Finland}
\affil[2]{Department of Industrial Engineering and Management, Aalto University School of Science, Espoo, Finland}
\affil[3]{Department of Experimental Psychology, University of Oxford, Oxford, UK}
\affil[4]{The Alan Turing Institute, London, UK}

\affil[*]{chandreyee.roy@aalto.fi}

\begin{abstract}
Humans are social animals and the interpersonal bonds formed between them are crucial for their development and well being in a society.
These relationships are usually structured into several layers (Dunbar's layers of friendship) depending on their significance in an individual's life with closest friends and family being the most important ones taking 
major part of their time and communication effort. However, we have little idea how the initiation and termination of these relationships occurs across the lifespan. To explore this, we analyse a national cellphone database to determine how and when changes in close relationships occur in the two genders.
In general, membership of this inner circle of intimate relationships is extremely stable, at least over a three-year period. However, around $1-4 \% $ of alters change every year, with the rate of change being higher among 17-21 year olds than older adults. Young adult females terminate more of their opposite-gender relationships, while older males are more persistent in trying to maintain relationships in decline. These results emphasise the variability in relationship dynamics across age and gender, and remind us that individual differences play an important role in the structure of social networks. Overall, our study provides a holistic understanding of the dynamic nature of relationships during the life-course of humans.

\end{abstract}

\begin{document}

\flushbottom
\maketitle
%
%
\thispagestyle{empty}


\section*{Introduction}
Human societies are built up out of personal social networks.
These networks evolve through formation of new bonds between pairs of individuals 
along with fall-outs of relationships between them.
We know relatively little about the rates of churn in networks and the dynamics of relationship change -- despite the fact that these are fundamental aspects of social networks. Previous research has shed some light on the temporal evolution of relationships that are manifestations of life-course events~\cite{roberts2011costs,bhattacharya2016sex}, geography\cite{bhattacharya2017absence}, structural changes in personal networks\cite{martin2006persistence} and patterns of mixing\cite{rivera2010dynamics}, among others \cite{bhattacharya2019social,pentland2015social,raeder2011predictors,miritello2013limited}. In this paper we examine the dynamics of relationship growth and decay with respect to ego age and gender using real world data from mobile phone records. We compare these turnovers with respect to different types of relationships. This gives us better understanding of the apparent competition between the needs for stability and diversity during the life-course of humans.
Human social networks have a layered structure, defined by the fractal structure of the Dunbar number \cite{dunbar2020structure}. This fractal structure forms a series of layers in the network that reflect the emotional closeness of the individuals involved 
\cite{zhou2005discrete,hill2003social,mac2016calling}. Studies on networks formed through social media, such as the Facebook and Twitter have shown layered structures similar to offline face-to-face networks~\cite{dunbar2015structure}, even in massive multiplayer online games~\cite{fuchs2014fractal}. The numbers 1.5, 5, 15, 50, and 150\cite{dunbar2021friends} cumulatively represent the number of intimate friendships, close friendships (or family members), best friends, good friends and friends, respectively. 
The individuals may lose or gain friends in each of these layers \cite{saramaki2014persistence,sutcliffe2012relationships} and the size of their egocentric networks has been observed to be higher for younger people and becomes more stable once they grow older\cite{bhattacharya2016sex,wrzus2013social}. The emotional closeness of an ego or focal individual with alters or individuals connected to the ego in each of these layers is inversely proportional to the number of alters in them, with the innermost layer consisting of the most intimate friendships and are the closest of all \cite{dunbar1998social, hill2003social,mok2007did}. 
This layer includes parents, siblings or close friends as well as romantic partners from the first layer and they are the set of people the ego would rely on for advice, help and emotional support. 
The ego is most likely to invest most of her or his social time on this group of alters (or close friends).

\begin{figure*}[ht]
    \centering
    \includegraphics[width=0.85\textwidth,height=3in]{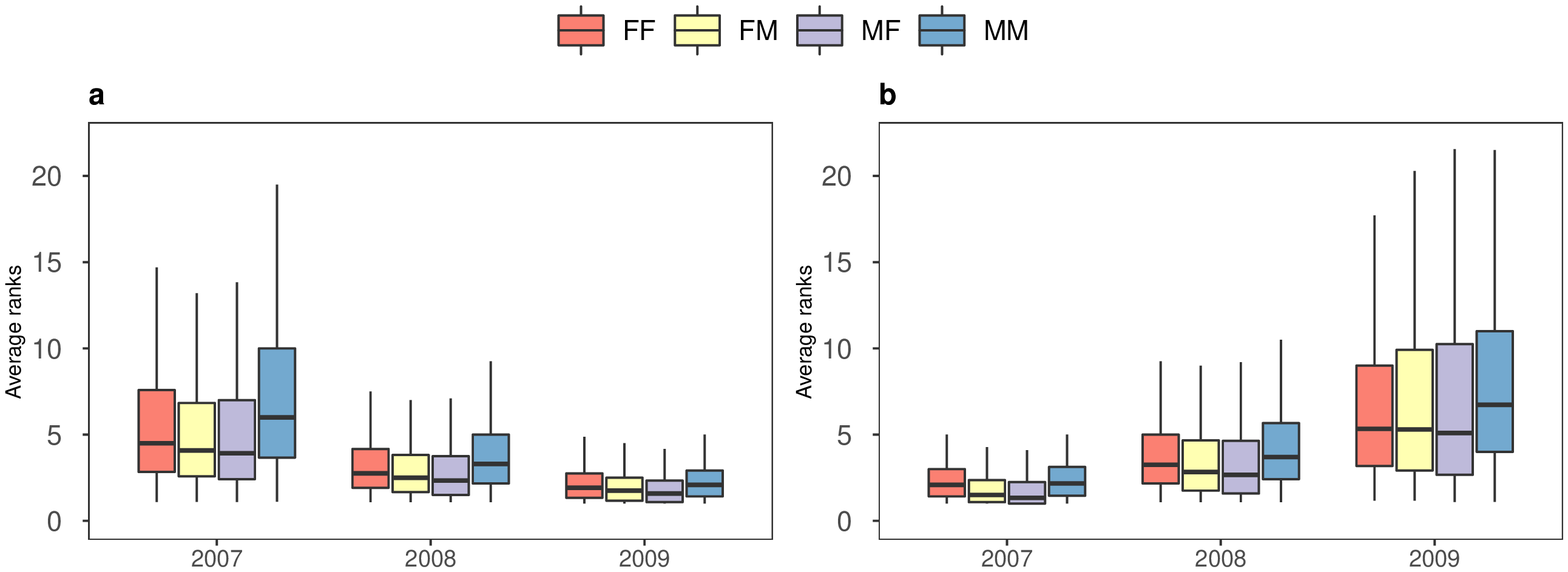}
    \caption{\textbf{Change in the rank of alters during the formation or decay of ego-alter relationships over a 3 year period}. The ranks have been calculated monthly, then averaged over 12 months for each year and a summary of their progressive increase/decrease over a span of three years are plotted in a boxplot for (a) the set of ego-alters forming close relationships and (b) another set of ego-alters decaying from close relationships.
    The black horizontal lines in the middle represent the median of the distribution. The box plot includes all the values within the range of the 25th and 75th percentile and the ends of the whiskers represent the maximum and the minimum ranks excluding outliers. The red box represents ranks of females and blue box represents the same for males in their respective same-gender egocentric networks. The yellow box represents the ranks of males in female egocentric networks and violet box represents the same for females in a male egocentric networks. We find that for all years in both cases, a male's rank among his male friends is consistently lower than other groups indicating that female friendships and opposite-gender friendships generally form stronger bonds than male friendships. 
    }
    \label{fig:ranks_3years}
\end{figure*}

In our study, we have considered the close relationships (``the support clique'') of the friendship network by studying their calling activities through mobile phone communication data of a particular service provider from a southern European country\cite{bhattacharya2019social}. This unique mobile phone dataset contains call detail records (CDRs) and demographic information of anonymized users that have remained loyal to the service provider for a period of three years \cite{onnela2007structure,blondel2015survey}. 
Analysing communication patterns through such 
mobile phone datasets are a valuable way to gain insight into the social networks of humans\cite{palchykov2012sex,miritello2013time} along with their behavioural patterns caused by seasonal and geographical changes\cite{monsivais2017seasonal,bhattacharya2017absence}, daily rhythms\cite{monsivais2017tracking,roy2021morningness,aledavood2018social} and mobility patterns\cite{gonzalez2008understanding,fudolig2021internal} among others~\cite{blondel2015survey,bhattacharya2019social}. A higher calling activity among a pair of individuals can be reasonably assumed to indicate a close relationship between them while a lower calling activity would indicate a weak tie. 
By the same token, we interpret a  decline in calling rate as indicating the termination of the relationship. If the decline is sudden (a step like change in call frequency), we interpret this as indicating a falling out. In other cases, the call frequency declines more gradually, suggesting a gradual loss of interest. 

\section*{Results}
\subsection*{Ranks of subscribers} 
In this study we have considered only those relationships between pairs of mobile phone users that form a close bond over a span of three years or decay from a close bond (see Methods). A pair of service subscribers is considered to have a close bond between them if both of them are in the top five rank of each others' social networks. Therefore, it is reasonable to assume that these two users fall within the first two Dunbar layers of their respective networks. A user with rank one corresponds to the topmost or highest rank. The ranks of users in all pairs have been calculated in their respective egocentric social networks and are grouped according to their gender (see the Methods section). The same-gender interactions are termed as female-female (FF) and male-male (MM) ego-alter friendships while the opposite-gender interactions are termed female-male (FM) and male-female (MF) ego-alter friendships, respectively. 

In Fig. \ref{fig:ranks_3years} we exhibit a summary of the ranks, in the form of a box plot, for males and females in their respective friend or partner's networks from the year 2007 to 2009. 
Formation of a close bond is represented by increased calling activity between the pair and a shift from lower to higher ranks ($ \leq 5$). This can also include relationships that have ranks between 2 to 5 in year 2007 and gradually increase to even higher values in the following years. 
The set of ego-alter pairs that show progressive increase in ranks in successive years are shown in of Fig. \ref{fig:ranks_3years}(a).
The colour red/blue represents ranks of females/males in their respective female/male partner's egocentric network. For opposite gender friendships, the colours yellow/violet represent ranks of males/females in the female/male partner's egocentric network. 
The ranks of the males in same gender friendships are observed to be noticeably lower than all the other groups for all three years indicating that males might not form as close bonds with their male counterparts when compared to female-female 
or even opposite-gender friendships. 

We have observed the same trend in Fig. \ref{fig:ranks_3years}(b) where we exhibit the ranks for only those ego-alter relationships that have decayed from a close bond by a gradual decrease in calling activity and a decrease from higher to lower ranks over a span of three years. These decaying relationships also include pairs whose ranks decrease to lower values but still remain $\leq 5$. The extent of a fall or rise in ranks in the years $2007 \xrightarrow[]{} 2008$ and $2008 \xrightarrow[]{} 2009$ is also greater in the case of male-male 
friendships, indicating that male friendships are not as close as opposite-gender or female-female 
friendships (see Supplementary Fig. 1).


\subsection*{Network turnover}

\begin{figure*}[!htp]
    \centering
    \includegraphics[width=\textwidth,height=3in]{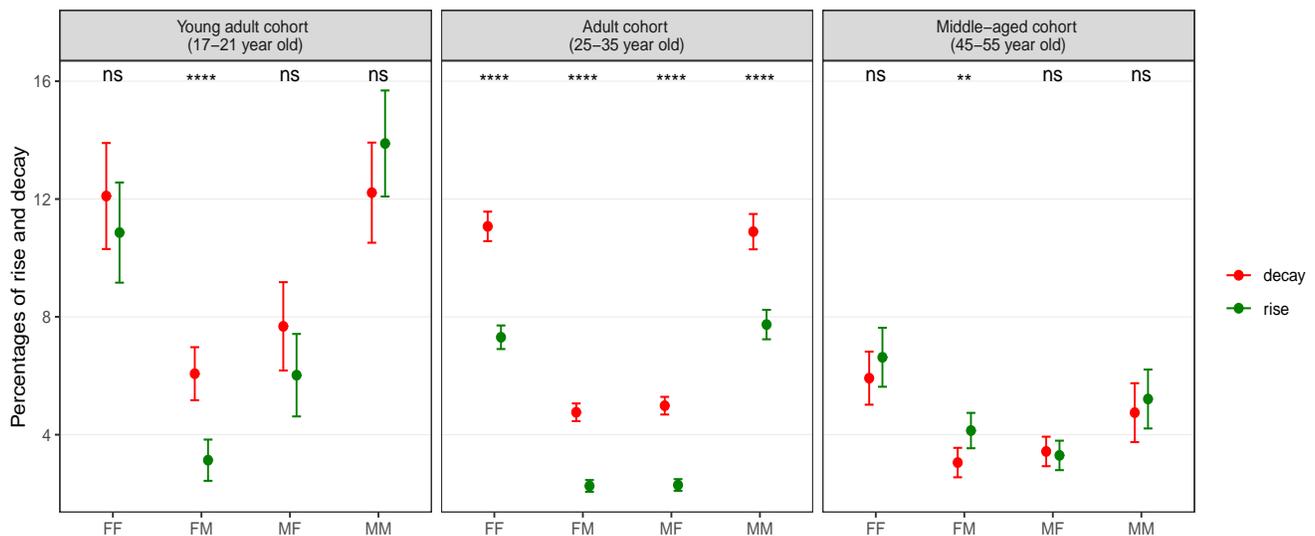}
    \caption{\textbf{Percentages of ego-alter relationships that have formed or decayed from a close bond among peers in three different age cohorts.} Total number of pairs in each group are given in Supplementary Table I. 
    The error bars are shown with $95\%$ confidence level and the p values from two proportions z-test are shown in the top of each panel to illustrate how significantly different the rise and the decay percentages are. We have used the following notation scheme to represent the significance levels of p values: not significant (ns) for $p>0.05$, and significant to the degree of  (*) for $p\leq 0.05$, (**) for $p \leq 0.01$, (***) for $p \leq 0.001$ and (****) for $p \leq 0.0001$. 
    Overall the percentage-wise rise and decay of friendship 
    for the young-adult 
    cohort are observed to be higher than those for the adult and middle-aged cohorts. We see a marked difference in the behaviour of female ego's relationships with their male alters with low values of rise and higher values of decay in the young-adult and adult groups. Moreover, we observe that young adult male egos have the comparatively same values of rise and decay with female alters but change their behaviour in the adult group. The percentage values corresponding to the graph are shown in Supplementary Table III.  
    }
    
    \label{fig:network_turnover_peers}
\end{figure*}

\begin{figure*}[!htp]
    \centering
    \includegraphics[width=\textwidth,height=3in]{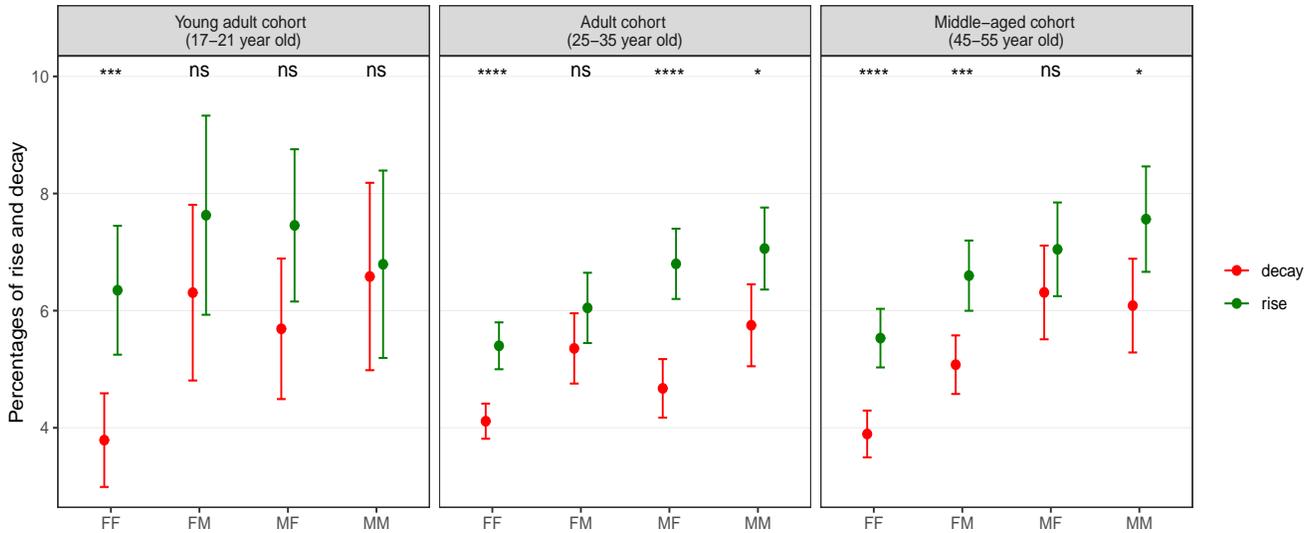}
    \caption{\textbf{Percentages of ego-alter relationships that have formed or decayed from a close bond among non-peers in three different age cohorts.}  Total number of pairs from each group are given in a corresponding Supplementary Table II. 
    The error bars are displayed with $95\%$ confidence level and the p values from two proportions z-test are shown in the top of each panel to illustrate how significantly different the rise and the decay percentages are using the same scheme as mentioned in Fig. \ref{fig:network_turnover_peers}. Here, we find that the percentages of rise are comparatively higher than percentages of decay for adult and middle-aged cohorts. Particularly, the FF group has a higher percentage of rise than decay in all age groups. The corresponding percentage values for this graph is shown in Supplementary Table IV.}
    
    \label{fig:network_turnover_nonpeers}
\end{figure*}

In order to investigate the turnover in friendship networks we
compared the calling activities between three different age groups: young adult cohorts (from 17 years to 21 years), adult cohorts (from 25 years to 35 years) and the middle-aged cohorts (from 45 to 55 years). We have considered these age groups as they broadly represent the important stages in a human life course.
When the age difference between an ego and an alter is less than or equal to 10 years, then
we deem it as a ``peer'' interaction and if the difference is between 20 to 40 years of age then we deem it to be a ``non-peer'' interaction.
The total number of pairs in each of the gender and age based groups have been enumerated in Supplementary Tables I and II for peer and non-peer interactions, respectively. We have then selected only those pairs that either form a close bond or decay from a close bond according to their calling activity and ranks as described in the methods section. In Fig. \ref{fig:network_turnover_peers} we have summarised the percentages for peer interactions and in Fig. \ref{fig:network_turnover_nonpeers} for non-peer interactions. 

Overall, the low values of the percentages in the case of formation and decay of close bonds in a three year period is an indicator of how stable these inner core relationships typically are. Nonetheless, there are some surprising details reflected in these data. The percentage of the relationship rise ($3.1 \pm 0.7 \%$) in opposite gender friendships for the 
female egos belonging to the young adult cohorts is almost half of their corresponding decay percentage ($6.1 \pm 0.9 \%$): in effect, they lose close opposite gender relationships over time without replacing them, although the rate at which this happens is slow (roughly $1\%$ of relationships per year). Conversely, the male egos in the young adult cohort have similar values for both the rise ($6.0 \pm 1.4\%$) and the decay ($7.7 \pm 1.5\%$) indicating that they make friends as easily as they lose them. The behaviour of the female egos in the young adult cohort is seen even in the adult cohort for opposite gender relationships. The rise percentages ($2.3 \pm 0.2\%$ for both male egos and female egos) in the adult cohorts are almost half of their decay percentages ($4.8 \pm 0.3\%$ and $5.0 \pm 0.3\%$ for female and male egos, respectively).
Therefore we find that the male and female egos behave similarly in the adult group but behave differently in the young adult group. 

For the same gender friendships, we observe that that the rise and decay percentages balance each other out in the young adult group for both female egos ($10.9 \pm 1.7\%$ and $12.1 \pm 1.8\%$ for the rise and decay, respectively) and the male egos ($13.9 \pm 1.8\%$ and $12.2 \pm 1.7\%$ for the rise and decay, respectively) with rates of turnover that are higher than in the adult and middle-aged cohorts. We also observe that they have a distinctively higher decay than rise particularly in the adult group with the percentages of the rise being $\{ 7.3 \pm 0.4\%$, $7.7 \pm 0.5\% \} $ and the percentage of the decay being $\{ 11.1 \pm 0.5\%$, $10.9 \pm 0.6\% \} $ for female and male egos, respectively. For the middle-age cohorts we observe that the percentages of the relationship rise and decay for all gender groups are in balance with values of the rise being $\{ 6.6 \pm 1.0\%$, $5.2 \pm 1.0\%$, $4.1 \pm 0.6\%$, $3.3 \pm 0.5\% \}$ and the decay being $\{5.9 \pm 0.9\%$, $4.7 \pm 1.0\%$, $3.1 \pm 0.5\%$, $3.4 \pm 0.5\% \}$ for FF, MM, FM, MF friendships, respectively, where the gender on the left is the ego and gender on the right is the alter. In other words, the turnover is high in the younger cohorts in all gender groups and becomes smaller as individuals age.

When the users interact with their non-peers (see Fig. \ref{fig:network_turnover_nonpeers}), we observe that the percentages of rise is higher than decay for all gender groups in the adult and middle-aged cohort indicating that they try to maintain their non-peer relationships. Particularly, we find that the percentage of rising relationships is significantly higher than decaying ones in FF group for all age cohorts. The percentages of rise are $\{ 6.3 \pm 1.1\%$, $5.4 \pm 0.4\%$, $5.5 \pm 0.5 \% \}$ and percentages of decay are $\{ 3.8 \pm 0.8\%$, $4.1 \pm 0.3\%$, $3.9 \pm 0.4 \% \}$ for young adult, adult and middle-aged cohorts, respectively, indicate that female egos form increasingly closer bonds with their older female alters in all age groups.

\subsection*{Temporal pattern of calling activity}

\begin{figure*}[!htp]
    \centering
    \includegraphics[width=\textwidth,height=6in]{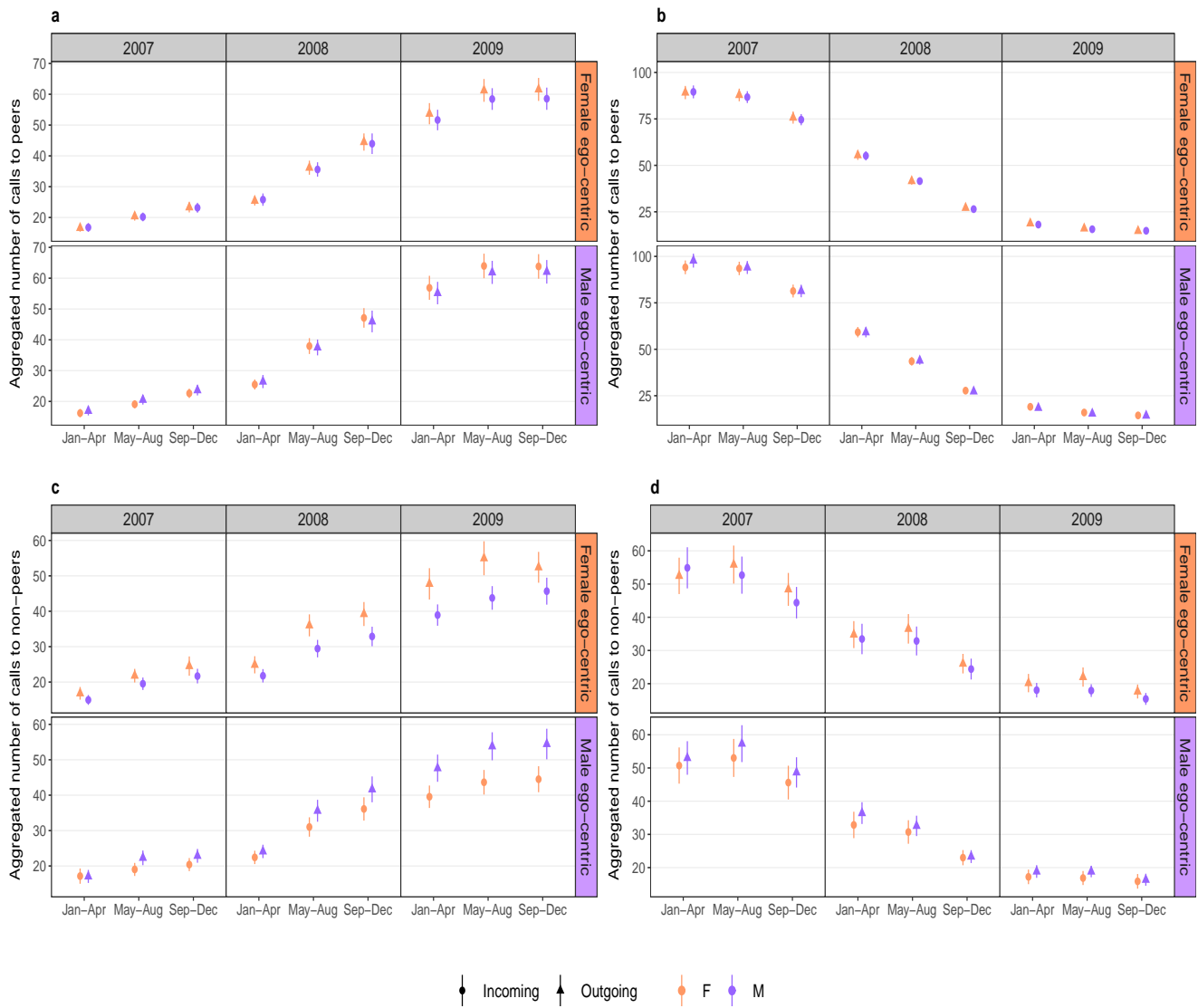}
    \caption{\textbf{Calling activities of adult (25 - 35 year old) cohorts between peers and non-peers and having interactions with their opposite genders.} The aggregated calls over four-month intervals over a span of three years have been calculated for the opposite-gender friendships for peers (having age differences of less than 10 years) exhibiting (a) formation of close bonds and (b) decaying of close bonds. We also show the aggregated calls of the young age cohorts calling their non-peers (having age differences of 20 to 40 years) for formation (c) and decay (d) of close bonds. 
    The orange/violet coloured panels on the right of each of the graphs represent the calls made in a female/male egocentric network. In all the graphs, filled circles/triangles represent incoming/outgoing calls and orange/violet coloured shapes indicate calls initiated by female/male egos in the egocentric networks. Therefore, an orange coloured triangle indicates calls made by females to males while violet coloured circles represent calls made by males to females in a female egocentric network. Similarly, orange coloured circles and violet coloured triangles indicate female-initiated and male-initiated calls in a male-egocentric network. 
    The female-initiated calls in both male and female egocentric networks in the year 2009 show systematically higher values than male-initiated calls in the case of formation of close relationships between peers 
    while there is not much difference visibly between them in decaying relationships. We also find that younger cohorts are more active than their older cohorts in terms of calling activities since they initiate distinctly more number of calls than their older counterparts when forming a close bond. For decaying relationships, we do not find much difference in the calling activities of both genders in female egocentric networks but in the male egocentric network we find that male initiated calls are slightly more female initiated ones.   
    }
    \label{fig:mf_young}
\end{figure*}
\begin{figure*}[!htp]
    \centering
    \includegraphics[width=\textwidth,height=6in]{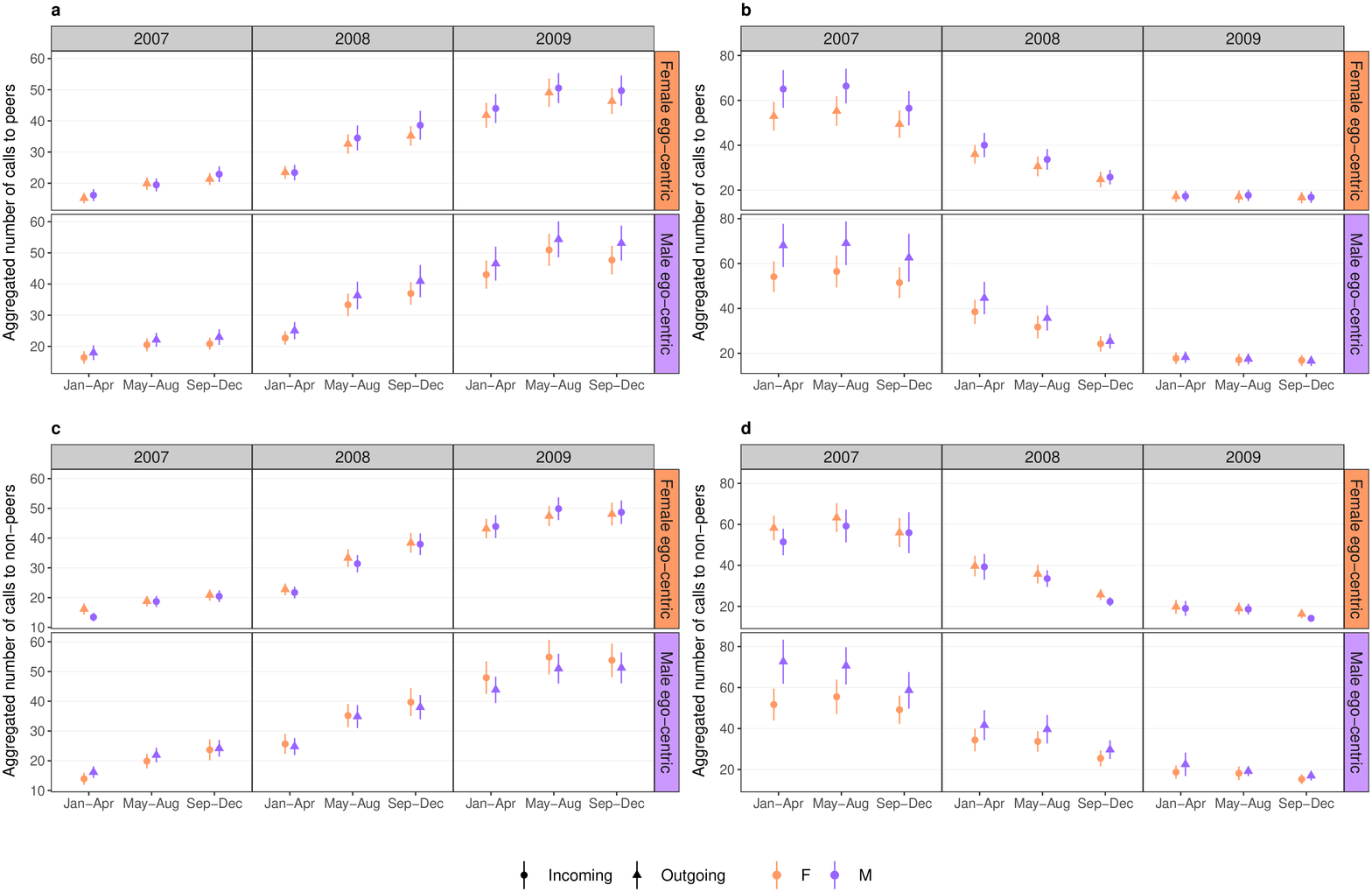}
    \caption{\textbf{Calling activities of middle-aged age (45 - 55 year old) cohorts between peers and non-peers and having interactions with their opposite genders.} The aggregated calling activities of middle-aged cohorts between their peers (having age differences of less than 10 years) has been shown in (a) and (b) along with non-peers (having age differences of 20 to 40 years) in (c) and (d). The plots on the left exhibit those relationships that form a close bond and the plots on the right exhibit decay of close bonds. The orange/violet coloured panels on right of each of the plots represent female/male egocentric network. The filled triangles/circles represent outgoing/incoming calls and orange/violet colour of the points represent calls initiated by females/males in their respective networks.
    We find that middle-aged individuals calling their peers have slightly higher male-initiated calls than females in both rising and decaying relationships. Their calling activity with non-peers show a consistent behaviour with the young age cohorts' interaction with non-peers in Fig. \ref{fig:mf_young}(c) where they are receiving slightly higher number of calls from the younger opposite gender individuals. However, we find that when relationship decays the males are making higher number of calls to the younger females.
    }
    \label{fig:mf_middleaged}
\end{figure*}

To explore these patterns in more detail in the adult cohort, Fig. \ref{fig:mf_young}(a) and (b) exhibit the calls made to peers aggregated in four month blocks for the formation and decay of close bonds, respectively. Fig. \ref{fig:mf_young}(c) and (d) represent the formation and decay of close bonds through calls made to non-peers. 
The decaying relationships for adult cohorts calling peers (see Fig. \ref{fig:mf_young}(b)) do not show much differences between calls initiated by males and those initiated by females. However, when they form a close bond, even though calls initiated by both genders are not significantly different from each other (i.e. p values > 0.05 obtained from paired t-tests), there is a striking tendency for calls made by females to their male counterparts to increase faster than those made by males, suggesting that females are more proactive than males in building opposite-gender relationships (see Fig. \ref{fig:mf_young}(a)). We also observe in Fig. \ref{fig:mf_young}(c) that both the genders call their older counterparts more often when they are trying to form a close bond with their elders (to a significant degree as p values < 0.05 obtained from paired t-tests), which suggests that younger individuals are more active in calling. In a decaying relationship we notice that there is not much difference in the number of calls initiated by both genders in a female egocentric network but in the male egocentric network the younger males are systematically calling the older females more than receiving calls from them. The corresponding graphs for the young adult cohort is displayed in the Supplementary Fig. 2 and we have observed similar calling patterns for all interactions except in the case of decaying relationships in female egocentric networks where male-initiated calls were higher than female-initiated calls.

In Fig. \ref{fig:mf_middleaged} we exhibit the calling activities of the middle-aged cohorts to their peers (Fig. \ref{fig:mf_middleaged}(a) and (b)) and their non-peers (Fig. \ref{fig:mf_middleaged}(c) and (d)). 
In the case of 
middle-aged cohorts calling their peers, we observe that in both formation and decay of close bonds (Fig. \ref{fig:mf_middleaged}(a) and (b), respectively) males initiated more calls than females, indicating that males take the more active part in trying to maintain a relationship. In the case of the middle-aged cohorts calling their non-peers during the formation of close bonds (Fig. \ref{fig:mf_middleaged}(c)), we observe that they are being called more by their younger counterparts of opposite gender as was the case in Fig. \ref{fig:mf_young}(c). For decaying relationships in Fig. \ref{fig:mf_middleaged}(d), we again observe that the older males call more than receive calls from their younger female counterparts. 
We also investigated gradual and abrupt fall of calling activity in decaying relationships by looking at the differences between their monthly calling activities. However, this method of differences did not procure any interesting results with respect to age and gender of the users involved. 


\section*{Discussion}
In this study we have analysed a population level mobile phone communication data from a Southern European country to investigate relationship changes in different phases of human life course. Since, all the users are from the same service provider that had a $20\%$ market share in the country, this dataset is able to capture only a proportion of the true support clique. However, even though the top five ranks may not be able to take into account the actual support clique of the users, it still provides us with a number of new insights into the dynamics of close relationships formed between the same-gender and opposite-gender individuals. Firstly, the lower ranks of the males in MM friendships when compared to FF friendships strongly suggest that females tend to have more intense friendships than males \cite{david2015women}. In general, the females tend to have fewer and more close relationships than males as 
can be seen through the comparison of the ranks in Fig. \ref{fig:ranks_3years}. A number of other gender differences may be noted. The median of the ranks in opposite-gender friendships has a systematic trend of females having higher ranks in their male counterpart's network than the male's rank in their male counterpart's network. This suggests that men tend to give more importance to the women (spouse, mother or romantic partner) in their lives than to their male friends. We also notice that the male's rank in their female partner's network is comparable to the rank of females in the same gender (FF) friendships. This is consistent with the finding that females tend to have two special friends in their core layer of friendship networks (a best friend who is usually a female and a romantic partner) while men tend to have only one special friend (either a romantic partner or a male friend, but rarely both at the same time). \cite{dunbar2021friends} 

Secondly, the overall low percentages for the formation/decay of relationships in the network as observed in Figures \ref{fig:network_turnover_peers} and \ref{fig:network_turnover_nonpeers} for mutual top 5 ranked alters that form the support clique of the egocentric network reflect a reassuring fact that close friendships mostly remain stable and rarely show fall-outs between them. It is known that this core layer of the network comprises parents, romantic partners and best friends and therefore, these important relationships should not be expected to change much over the lifespan of an individual. 

Thirdly, the percentages of the formation/decay of close relationships are observed to be higher in the young adult and adult cohorts than the middle-aged cohorts indicating more stable friendships in the latter. The young adults in this age group are usually more mobile, socially and economically, and this may be expected to impact on the stability of their relationships.
We notice that, particularly for opposite gender friendships, the females have a lower rise than decay, indicating that they may be more meticulous in choosing male friends. \cite{pawlwski1999impact} The same 
pattern is not observed in the case of young adult male egos. This could also be due to the fact that guardianship of the family over young adult females tend to be exercised over a longer period than young adult males in the society. 

Fourthly, marked differences in the percentages of the relationship rise and decay in the case of 25-35 year old adults for opposite gender friendships for both the male and female egos likely reflects the impact of reproduction: in this population, 
the mean age at marriage at the time the phone data were sampled was 29 years, with first reproduction following soon afterwards.\cite{coleman2013partnership}
Moving out of the parental home to establish an independent home life, combined with the arrival of children, marks a major social transition and upheaval \cite{rozer2015romantic,johnson1982couple}. 
We also observe that the percentages of the rise and decay eventually balance out in the middle-aged cohorts thus indicating that this age group has 
reached stability in their lives. 

Finally, we observe from Fig. \ref{fig:network_turnover_nonpeers} that most relationships in all gender and age groups with non-peer interactions 
have higher 
percentages of rise than decay and in particular for the FF group, in which 
we see a consistent higher rise than decay in all age cohorts.
This could be because the young adult group start moving out of their parental home resulting in increased calling activity with their parents. The adult cohort also show 
higher percentages of rise than decay. This could be the result of their hitting the reproductive stage, which leads them to retain only those non-peer relationships (like aunties and others) that help them in tasks related to child rearing among others. Another reason for the rise in the middle-aged group could also be 
due to the grandmothering effect\cite{hawkes2004grandmother} that comes into play when the mothers of adults start taking an interest in their children's lives for child rearing reasons\cite{sear2000maternal}.


Additionally, we have observed differences in the calling activity patterns of the opposite-gender peer and non-peer interactions among the adult and middle-age cohorts while forming and decaying a relationship. Our analyses show that after a close bond is formed between opposite gender peers in adult cohorts, it is the females that systematically tend to call their male counterparts more often.
However, when the middle-age cohorts interact with their peers, then the males call their female counterparts more for both the cases when relationships are being formed and when they are decaying. Since we do not expect these kinds of relationships to happen between siblings, this could probably represent romantic or platonic friendships. In addition, we observe 
a distinctly higher call rate by the adult cohorts to their older counterparts when the relationships are being formed. The formation of relationships in this group may be indicative of those being formed between an individual and their in-laws and could represent that the adult cohorts take more effort in trying to establish a relationship with their older counterparts since younger age groups are generally more active technologically than older age groups.

In conclusion, our mobile phone data analyses of a large population establish that the inner core Dunbar layer of friendships are stable with very few relationships moving out or entering this core. During the younger stages of adult life people mostly lose friends, and particularly the young adults have slightly more unstable friendships as compared to adults and this is mostly because at this stage of their lives they make a lot of friends and lose a lot of them as well. However, ultimately the friendships become more stable in the middle-age after people settle down with families and stable jobs. Additionally, young females tend to be picky while choosing male friends, but after they form a close bond, they tend to be slightly more active in those relationships.

\section*{Methods}
\subsection*{Study data}
The dataset used in this study was obtained from a southern European service provider consisting of mobile phone call detail records (CDRs) of the service users for a period of three years from January, 2007 to December, 2009. The call details of the users were anonymised by the service provider themselves by attaching a unique identifier for each of the users such that the privacy of the customers remain protected and cannot be traced back to the individuals themselves. The CDRs include a list of the caller and callee with time stamps and dates. The call durations for each call are provided only for the year 2007. The users' age and gender along with the postal code 
are also listed in the dataset. A total of 644,170 remained loyal to the service provider for all three years with a total of 1,143,718 unique links between them.


\subsection*{Data analyses}
\begin{figure}
\begin{center}
    \begin{tikzpicture}[>=latex]
  \node [rect] (base) {$L_{2007}^1$} ;
  \node [rect, right =10em of base] (A) {$L_{2009}^1$}
  edge [from] node[above =0pt, name=e1] {Formation  } (base)
  edge [from] node[below =0pt, name=e2] {Decay} (base);
  \draw[->] (e1) ;
\draw[->] (e2)  ;
  \node [rect, below =5em of base] (B) {$L_{2007}^n$}
  edge [towards] node[sloped, above =0pt, name=e5, pos= 0.23] {Formation} (A);
 \draw[->] (e5) ; 
  \node [rect, right=10 em of B] (C) {$L_{2009}^n$}
  edge [from] node[sloped,anchor = right, above =0pt, name=e3, pos= 0.25] {Decay} (base)
edge [from] node[above=0pt,name=e4] {Not considered} (B);
\draw[->] (e3) ;
\draw[->] (e4)  ;
\end{tikzpicture} 

\end{center}   
\caption{\textbf{The relationships that have been considered for analyses.} All pairs that were mutually in top 5 ranks in the year 2007 (represented by $L_{2007}^1$) and after a rise or decay in call activity still remain in top 5 in the year 2009 ($L_{2009}^1$) along with pairs whose ranks become lower than 5 ($L_{2007}^1 \longrightarrow L_{2009}^n$: decay where \textit{n} in the superscript represents any rank lower than 5) or rise from a lower rank to a higher one ($L_{2007}^n \longrightarrow L_{2009}^1$: formation) have been considered. Only pairs that remain in lower ranks have not been considered. The arrows have been labelled accordingly to show the change in the ranks. Relationships having users with ranks lower than 5 in 2007 and remain low in 2009 have not been considered.}
\label{fig:diagram}
\end{figure}
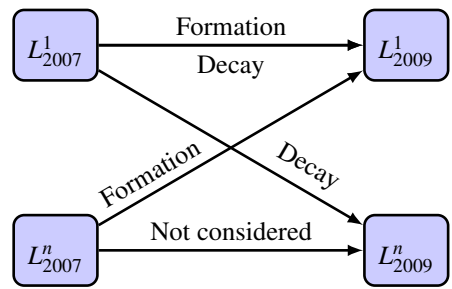

The rank of a user in another user's network has been calculated in the following way. We consider all the alters that the user has called or received calls from on a monthly basis and have ordered them according to the total number of Incoming/Outgoing calls in a decreasing order. Alters having the highest number are ranked one and so on. The ranks are then calculated for each of the 36 months being considered and then averaged on a yearly basis. Next, we consider only those links with users having top ranks between 1 to 5 mutually in each others networks in the year 2007 and 2009. This is because we intend to study formation or decay of very close relationships that form the "support clique" of the Dunbar layers. Thus, any pair that starts off with a mutual top 5 rank and then decays over the span of three years to lower ranks and any pair that starts of at any rank but ultimately ends with top 5 rank in the year 2009 have been considered. In such a scenario, if we represent all pairs having top 5 ranks as $L^1$ and have ranks lower than 5 as $L^n$ (where \textit{n} in the superscript represents any rank lower than 5) then we consider only the changes in ranks of the relationships as shown by arrows in a schematic diagram in Fig. \ref{fig:diagram}. The ranks could still be in top 5 even though the relationship decays or rises from a top rank to an even higher rank. The only type of relationship that has not been considered are the ones that go from $L_{2007}^n  \longrightarrow L_{2009}^n$. Keeping this in mind we have a total of 178,592 links between 296,051 users with 44,115 female-female, 
106,570 female-male or male-female 
and 27,907 male-male 
friendships considered for analysis and the age distribution of the users is shown in Supplementary Fig. 3. Furthermore, the interaction between users is deemed as a peer relationship if the age gap between them is less than or equal to 10 years and non-peer if it is between 20 to 40 years. 

Next, we define a formation of a close relationship to occur between two pairs of users if the calling activity between them increases and yearly averaged ranks for both the users increases each year. On the other hand, a decay in relationship is defined to occur if calling activity decreases and yearly averaged rank decreases each year. This filtering of the data leads to 8,637 pairs of users who end up forming a close relationship in 2009 and 10,310 pairs whose relationship decayed over three years. The age distribution of alters of the egos considered in the three different categories (young adults between 17-21 years, adults between 25-35 years and middle-aged between 45-55 years of age) are shown in Supplementary Figs. 4 and 5 for peer and non-peer interactions among the opposite gender, respectively.

\section*{Acknowledgments}
C.R., K.B., and K.K. acknowledge support from EU HORIZON 2020 INFRAIA-1-2014-2015 program project (SoBigData) No. 654024 and INFRAIA-2019-1 (SoBigData++) No. 871042 as well as NordForsk Programme for Interdisciplinary Research project "The Network Dynamics of Ethnic Integration". KK also acknowledges the Visiting Fellowship at The Alan Turing Institute, UK.

\section*{Data Availability}
The datasets generated during and/or analysed during the current study are not publicly available due to a signed NDA but are available from the corresponding author on reasonable request.

\bibliography{sample}

\begin{thebibliography}{10}
\urlstyle{rm}
\expandafter\ifx\csname url\endcsname\relax
  \def\url#1{\texttt{#1}}\fi
\expandafter\ifx\csname urlprefix\endcsname\relax\def\urlprefix{URL }\fi
\expandafter\ifx\csname doiprefix\endcsname\relax\def\doiprefix{DOI: }\fi
\providecommand{\bibinfo}[2]{#2}
\providecommand{\eprint}[2][]{\url{#2}}

\bibitem{roberts2011costs}
\bibinfo{author}{Roberts, S.~G.} \& \bibinfo{author}{Dunbar, R. I.~M.}
\newblock \bibinfo{journal}{\bibinfo{title}{The costs of family and friends: an
  18-month longitudinal study of relationship maintenance and decay}}.
\newblock {\emph{\JournalTitle{Evolution and Human Behavior}}}
  \textbf{\bibinfo{volume}{32}}, \bibinfo{pages}{186--197}
  (\bibinfo{year}{2011}).

\bibitem{bhattacharya2016sex}
\bibinfo{author}{Bhattacharya, K.}, \bibinfo{author}{Ghosh, A.},
  \bibinfo{author}{Monsivais, D.}, \bibinfo{author}{Dunbar, R. I.~M.} \&
  \bibinfo{author}{Kaski, K.}
\newblock \bibinfo{journal}{\bibinfo{title}{Sex differences in social focus
  across the life cycle in humans}}.
\newblock {\emph{\JournalTitle{Royal Society open science}}}
  \textbf{\bibinfo{volume}{3}}, \bibinfo{pages}{160097} (\bibinfo{year}{2016}).

\bibitem{bhattacharya2017absence}
\bibinfo{author}{Bhattacharya, K.}, \bibinfo{author}{Ghosh, A.},
  \bibinfo{author}{Monsivais, D.}, \bibinfo{author}{Dunbar, R. I.~M.} \&
  \bibinfo{author}{Kaski, K.}
\newblock \bibinfo{journal}{\bibinfo{title}{Absence makes the heart grow
  fonder: social compensation when failure to interact risks weakening a
  relationship}}.
\newblock {\emph{\JournalTitle{EPJ data science}}}
  \textbf{\bibinfo{volume}{6}}, \bibinfo{pages}{1--10} (\bibinfo{year}{2017}).

\bibitem{martin2006persistence}
\bibinfo{author}{Martin, J.~L.} \& \bibinfo{author}{Yeung, K.-T.}
\newblock \bibinfo{journal}{\bibinfo{title}{Persistence of close personal ties
  over a 12-year period}}.
\newblock {\emph{\JournalTitle{Social Networks}}}
  \textbf{\bibinfo{volume}{28}}, \bibinfo{pages}{331--362}
  (\bibinfo{year}{2006}).

\bibitem{rivera2010dynamics}
\bibinfo{author}{Rivera, M.~T.}, \bibinfo{author}{Soderstrom, S.~B.} \&
  \bibinfo{author}{Uzzi, B.}
\newblock \bibinfo{journal}{\bibinfo{title}{Dynamics of dyads in social
  networks: Assortative, relational, and proximity mechanisms}}.
\newblock {\emph{\JournalTitle{annual Review of Sociology}}}
  \textbf{\bibinfo{volume}{36}}, \bibinfo{pages}{91--115}
  (\bibinfo{year}{2010}).

\bibitem{bhattacharya2019social}
\bibinfo{author}{Bhattacharya, K.} \& \bibinfo{author}{Kaski, K.}
\newblock \bibinfo{journal}{\bibinfo{title}{Social physics: uncovering human
  behaviour from communication}}.
\newblock {\emph{\JournalTitle{Advances in Physics: X}}}
  \textbf{\bibinfo{volume}{4}}, \bibinfo{pages}{1527723}
  (\bibinfo{year}{2019}).

\bibitem{pentland2015social}
\bibinfo{author}{Pentland, A.}
\newblock \emph{\bibinfo{title}{Social Physics: How social networks can make us
  smarter}} (\bibinfo{publisher}{Penguin}, \bibinfo{year}{2015}).

\bibitem{raeder2011predictors}
\bibinfo{author}{Raeder, T.}, \bibinfo{author}{Lizardo, O.},
  \bibinfo{author}{Hachen, D.} \& \bibinfo{author}{Chawla, N.~V.}
\newblock \bibinfo{journal}{\bibinfo{title}{Predictors of short-term decay of
  cell phone contacts in a large scale communication network}}.
\newblock {\emph{\JournalTitle{Social Networks}}}
  \textbf{\bibinfo{volume}{33}}, \bibinfo{pages}{245--257}
  (\bibinfo{year}{2011}).

\bibitem{miritello2013limited}
\bibinfo{author}{Miritello, G.}, \bibinfo{author}{Lara, R.},
  \bibinfo{author}{Cebrian, M.} \& \bibinfo{author}{Moro, E.}
\newblock \bibinfo{journal}{\bibinfo{title}{Limited communication capacity
  unveils strategies for human interaction}}.
\newblock {\emph{\JournalTitle{Scientific reports}}}
  \textbf{\bibinfo{volume}{3}}, \bibinfo{pages}{1--7} (\bibinfo{year}{2013}).

\bibitem{dunbar2020structure}
\bibinfo{author}{Dunbar, R. I.~M.}
\newblock \bibinfo{journal}{\bibinfo{title}{Structure and function in human and
  primate social networks: implications for diffusion, network stability and
  health}}.
\newblock {\emph{\JournalTitle{Proceedings of the Royal Society A}}}
  \textbf{\bibinfo{volume}{476}}, \bibinfo{pages}{20200446}
  (\bibinfo{year}{2020}).

\bibitem{zhou2005discrete}
\bibinfo{author}{Zhou, W.-X.}, \bibinfo{author}{Sornette, D.},
  \bibinfo{author}{Hill, R.~A.} \& \bibinfo{author}{Dunbar, R. I.~M.}
\newblock \bibinfo{journal}{\bibinfo{title}{Discrete hierarchical organization
  of social group sizes}}.
\newblock {\emph{\JournalTitle{Proceedings of the Royal Society B: Biological
  Sciences}}} \textbf{\bibinfo{volume}{272}}, \bibinfo{pages}{439--444}
  (\bibinfo{year}{2005}).

\bibitem{hill2003social}
\bibinfo{author}{Hill, R.~A.} \& \bibinfo{author}{Dunbar, R. I.~M.}
\newblock \bibinfo{journal}{\bibinfo{title}{Social network size in humans}}.
\newblock {\emph{\JournalTitle{Human nature}}} \textbf{\bibinfo{volume}{14}},
  \bibinfo{pages}{53--72} (\bibinfo{year}{2003}).

\bibitem{mac2016calling}
\bibinfo{author}{Mac~Carron, P.}, \bibinfo{author}{Kaski, K.} \&
  \bibinfo{author}{Dunbar, R. I.~M.}
\newblock \bibinfo{journal}{\bibinfo{title}{Calling dunbar's numbers}}.
\newblock {\emph{\JournalTitle{Social Networks}}}
  \textbf{\bibinfo{volume}{47}}, \bibinfo{pages}{151--155}
  (\bibinfo{year}{2016}).

\bibitem{dunbar2015structure}
\bibinfo{author}{Dunbar, R. I.~M.}, \bibinfo{author}{Arnaboldi, V.},
  \bibinfo{author}{Conti, M.} \& \bibinfo{author}{Passarella, A.}
\newblock \bibinfo{journal}{\bibinfo{title}{The structure of online social
  networks mirrors those in the offline world}}.
\newblock {\emph{\JournalTitle{Social networks}}}
  \textbf{\bibinfo{volume}{43}}, \bibinfo{pages}{39--47}
  (\bibinfo{year}{2015}).

\bibitem{fuchs2014fractal}
\bibinfo{author}{Fuchs, B.}, \bibinfo{author}{Sornette, D.} \&
  \bibinfo{author}{Thurner, S.}
\newblock \bibinfo{journal}{\bibinfo{title}{Fractal multi-level organisation of
  human groups in a virtual world}}.
\newblock {\emph{\JournalTitle{Scientific reports}}}
  \textbf{\bibinfo{volume}{4}}, \bibinfo{pages}{1--6} (\bibinfo{year}{2014}).

\bibitem{dunbar2021friends}
\bibinfo{author}{Dunbar, R. I.~M.}
\newblock \emph{\bibinfo{title}{Friends: Understanding the Power of Our Most
  Important Relationships}} (\bibinfo{publisher}{Little, Brown Book Group},
  \bibinfo{year}{2021}).

\bibitem{saramaki2014persistence}
\bibinfo{author}{Saram{\"a}ki, J.} \emph{et~al.}
\newblock \bibinfo{journal}{\bibinfo{title}{Persistence of social signatures in
  human communication}}.
\newblock {\emph{\JournalTitle{Proceedings of the National Academy of
  Sciences}}} \textbf{\bibinfo{volume}{111}}, \bibinfo{pages}{942--947}
  (\bibinfo{year}{2014}).

\bibitem{sutcliffe2012relationships}
\bibinfo{author}{Sutcliffe, A.}, \bibinfo{author}{Dunbar, R. I.~M.},
  \bibinfo{author}{Binder, J.} \& \bibinfo{author}{Arrow, H.}
\newblock \bibinfo{journal}{\bibinfo{title}{Relationships and the social brain:
  integrating psychological and evolutionary perspectives}}.
\newblock {\emph{\JournalTitle{British journal of psychology}}}
  \textbf{\bibinfo{volume}{103}}, \bibinfo{pages}{149--168}
  (\bibinfo{year}{2012}).

\bibitem{wrzus2013social}
\bibinfo{author}{Wrzus, C.}, \bibinfo{author}{H{\"a}nel, M.},
  \bibinfo{author}{Wagner, J.} \& \bibinfo{author}{Neyer, F.~J.}
\newblock \bibinfo{journal}{\bibinfo{title}{Social network changes and life
  events across the life span: a meta-analysis.}}
\newblock {\emph{\JournalTitle{Psychological bulletin}}}
  \textbf{\bibinfo{volume}{139}}, \bibinfo{pages}{53} (\bibinfo{year}{2013}).

\bibitem{dunbar1998social}
\bibinfo{author}{Dunbar, R. I.~M.}
\newblock \bibinfo{journal}{\bibinfo{title}{The social brain hypothesis}}.
\newblock {\emph{\JournalTitle{Evolutionary Anthropology: Issues, News, and
  Reviews: Issues, News, and Reviews}}} \textbf{\bibinfo{volume}{6}},
  \bibinfo{pages}{178--190} (\bibinfo{year}{1998}).

\bibitem{mok2007did}
\bibinfo{author}{Mok, D.}, \bibinfo{author}{Wellman, B.} \emph{et~al.}
\newblock \bibinfo{journal}{\bibinfo{title}{Did distance matter before the
  internet?: Interpersonal contact and support in the 1970s}}.
\newblock {\emph{\JournalTitle{Social networks}}}
  \textbf{\bibinfo{volume}{29}}, \bibinfo{pages}{430--461}
  (\bibinfo{year}{2007}).

\bibitem{onnela2007structure}
\bibinfo{author}{Onnela, J.-P.} \emph{et~al.}
\newblock \bibinfo{journal}{\bibinfo{title}{Structure and tie strengths in
  mobile communication networks}}.
\newblock {\emph{\JournalTitle{Proceedings of the national academy of
  sciences}}} \textbf{\bibinfo{volume}{104}}, \bibinfo{pages}{7332--7336}
  (\bibinfo{year}{2007}).

\bibitem{blondel2015survey}
\bibinfo{author}{Blondel, V.~D.}, \bibinfo{author}{Decuyper, A.} \&
  \bibinfo{author}{Krings, G.}
\newblock \bibinfo{journal}{\bibinfo{title}{A survey of results on mobile phone
  datasets analysis}}.
\newblock {\emph{\JournalTitle{EPJ data science}}}
  \textbf{\bibinfo{volume}{4}}, \bibinfo{pages}{10} (\bibinfo{year}{2015}).

\bibitem{palchykov2012sex}
\bibinfo{author}{Palchykov, V.}, \bibinfo{author}{Kaski, K.},
  \bibinfo{author}{Kert{\'e}sz, J.}, \bibinfo{author}{Barab{\'a}si, A.-L.} \&
  \bibinfo{author}{Dunbar, R. I.~M.}
\newblock \bibinfo{journal}{\bibinfo{title}{Sex differences in intimate
  relationships}}.
\newblock {\emph{\JournalTitle{Scientific reports}}}
  \textbf{\bibinfo{volume}{2}}, \bibinfo{pages}{1--5} (\bibinfo{year}{2012}).

\bibitem{miritello2013time}
\bibinfo{author}{Miritello, G.} \emph{et~al.}
\newblock \bibinfo{journal}{\bibinfo{title}{Time as a limited resource:
  Communication strategy in mobile phone networks}}.
\newblock {\emph{\JournalTitle{Social networks}}}
  \textbf{\bibinfo{volume}{35}}, \bibinfo{pages}{89--95}
  (\bibinfo{year}{2013}).

\bibitem{monsivais2017seasonal}
\bibinfo{author}{Monsivais, D.}, \bibinfo{author}{Bhattacharya, K.},
  \bibinfo{author}{Ghosh, A.}, \bibinfo{author}{Dunbar, R. I.~M.} \&
  \bibinfo{author}{Kaski, K.}
\newblock \bibinfo{journal}{\bibinfo{title}{Seasonal and geographical impact on
  human resting periods}}.
\newblock {\emph{\JournalTitle{Scientific reports}}}
  \textbf{\bibinfo{volume}{7}}, \bibinfo{pages}{1--10} (\bibinfo{year}{2017}).

\bibitem{monsivais2017tracking}
\bibinfo{author}{Monsivais, D.}, \bibinfo{author}{Ghosh, A.},
  \bibinfo{author}{Bhattacharya, K.}, \bibinfo{author}{Dunbar, R. I.~M.} \&
  \bibinfo{author}{Kaski, K.}
\newblock \bibinfo{journal}{\bibinfo{title}{Tracking urban human activity from
  mobile phone calling patterns}}.
\newblock {\emph{\JournalTitle{PLoS computational biology}}}
  \textbf{\bibinfo{volume}{13}}, \bibinfo{pages}{e1005824}
  (\bibinfo{year}{2017}).

\bibitem{roy2021morningness}
\bibinfo{author}{Roy, C.}, \bibinfo{author}{Monsivais, D.},
  \bibinfo{author}{Bhattacharya, K.}, \bibinfo{author}{Dunbar, R. I.~M.} \&
  \bibinfo{author}{Kaski, K.}
\newblock \bibinfo{journal}{\bibinfo{title}{Morningness-eveningness assessment
  from mobile phone communication analysis}}.
\newblock {\emph{\JournalTitle{bioRxiv}}}  (\bibinfo{year}{2021}).

\bibitem{aledavood2018social}
\bibinfo{author}{Aledavood, T.}, \bibinfo{author}{Lehmann, S.} \&
  \bibinfo{author}{Saram{\"a}ki, J.}
\newblock \bibinfo{journal}{\bibinfo{title}{Social network differences of
  chronotypes identified from mobile phone data}}.
\newblock {\emph{\JournalTitle{EPJ Data Science}}}
  \textbf{\bibinfo{volume}{7}}, \bibinfo{pages}{46} (\bibinfo{year}{2018}).

\bibitem{gonzalez2008understanding}
\bibinfo{author}{Gonzalez, M.~C.}, \bibinfo{author}{Hidalgo, C.~A.} \&
  \bibinfo{author}{Barabasi, A.-L.}
\newblock \bibinfo{journal}{\bibinfo{title}{Understanding individual human
  mobility patterns}}.
\newblock {\emph{\JournalTitle{nature}}} \textbf{\bibinfo{volume}{453}},
  \bibinfo{pages}{779--782} (\bibinfo{year}{2008}).

\bibitem{fudolig2021internal}
\bibinfo{author}{Fudolig, M. I.~D.}, \bibinfo{author}{Monsivais, D.},
  \bibinfo{author}{Bhattacharya, K.}, \bibinfo{author}{Jo, H.-H.} \&
  \bibinfo{author}{Kaski, K.}
\newblock \bibinfo{journal}{\bibinfo{title}{Internal migration and mobile
  communication patterns among pairs with strong ties}}.
\newblock {\emph{\JournalTitle{EPJ Data Science}}}
  \textbf{\bibinfo{volume}{10}}, \bibinfo{pages}{1--21} (\bibinfo{year}{2021}).

\bibitem{david2015women}
\bibinfo{author}{David-Barrett, T.} \emph{et~al.}
\newblock \bibinfo{journal}{\bibinfo{title}{Women favour dyadic relationships,
  but men prefer clubs: cross-cultural evidence from social networking}}.
\newblock {\emph{\JournalTitle{PloS one}}} \textbf{\bibinfo{volume}{10}},
  \bibinfo{pages}{e0118329} (\bibinfo{year}{2015}).

\bibitem{pawlwski1999impact}
\bibinfo{author}{Paw{\l}wski, B.} \& \bibinfo{author}{Dunbar, R.~I.}
\newblock \bibinfo{journal}{\bibinfo{title}{Impact of market value on human
  mate choice decisions}}.
\newblock {\emph{\JournalTitle{Proceedings of the Royal Society of London.
  Series B: Biological Sciences}}} \textbf{\bibinfo{volume}{266}},
  \bibinfo{pages}{281--285} (\bibinfo{year}{1999}).

\bibitem{coleman2013partnership}
\bibinfo{author}{Coleman, D.}
\newblock \bibinfo{journal}{\bibinfo{title}{Partnership in europe; its variety,
  trends and dissolution}}.
\newblock {\emph{\JournalTitle{Finnish Yearbook of Population Research}}}
  \textbf{\bibinfo{volume}{48}}, \bibinfo{pages}{5--49} (\bibinfo{year}{2013}).

\bibitem{rozer2015romantic}
\bibinfo{author}{R{\"o}zer, J.~J.}, \bibinfo{author}{Mollenhorst, G.} \&
  \bibinfo{author}{Volker, B.}
\newblock \bibinfo{journal}{\bibinfo{title}{Romantic relationship formation,
  maintenance and changes in personal networks}}.
\newblock {\emph{\JournalTitle{Advances in life course research}}}
  \textbf{\bibinfo{volume}{23}}, \bibinfo{pages}{86--97}
  (\bibinfo{year}{2015}).

\bibitem{johnson1982couple}
\bibinfo{author}{Johnson, M.~P.} \& \bibinfo{author}{Leslie, L.}
\newblock \bibinfo{journal}{\bibinfo{title}{Couple involvement and network
  structure: A test of the dyadic withdrawal hypothesis}}.
\newblock {\emph{\JournalTitle{Social psychology quarterly}}}
  \bibinfo{pages}{34--43} (\bibinfo{year}{1982}).

\bibitem{hawkes2004grandmother}
\bibinfo{author}{Hawkes, K.}
\newblock \bibinfo{journal}{\bibinfo{title}{The grandmother effect}}.
\newblock {\emph{\JournalTitle{Nature}}} \textbf{\bibinfo{volume}{428}},
  \bibinfo{pages}{128--129} (\bibinfo{year}{2004}).

\bibitem{sear2000maternal}
\bibinfo{author}{Sear, R.}, \bibinfo{author}{Mace, R.} \&
  \bibinfo{author}{McGregor, I.~A.}
\newblock \bibinfo{journal}{\bibinfo{title}{Maternal grandmothers improve
  nutritional status and survival of children in rural gambia}}.
\newblock {\emph{\JournalTitle{Proceedings of the Royal Society of London.
  Series B: Biological Sciences}}} \textbf{\bibinfo{volume}{267}},
  \bibinfo{pages}{1641--1647} (\bibinfo{year}{2000}).

\end{thebibliography}


\section*{Author contributions statement}
All authors contributed in developing the scope of the paper. C.R. developed the numerical code along with figures and tables and wrote the first draft of the text. All authors contributed in interpretation, discussion, analysis of results and text improvement. All authors reviewed the manuscript.  


\section*{Competing interests}
The authors declare no competing interests.



\end{document}